\documentclass[dvipdfm,12pt,a4paper]{mdpi} 

\journal{Sensors}
\sysname{sensors}
\issn{1424-8220}
\logo{sensors}
\logosize{32}
\definecolor{logocolor}{rgb}{0,0.502,0.502}
\pagerange{5989-6007}
\doinum{10.3390/s90805989}
\pubyear{2009}
\pubvolume{9}

\usepackage{graphicx}
\usepackage{subfigure}
\usepackage{algorithmic}
\usepackage{algorithm}
\usepackage{amsfonts}
\usepackage{times}
\usepackage{lscape}
\usepackage[sort&compress]{natbib}
\usepackage{hyperref}
\usepackage{hypernat}
\hypersetup{bookmarks=true, bookmarksopen=true, unicode=false,
pdftoolbar=true, pdfmenubar=true, pdffitwindow=true,
pdfnewwindow=true, colorlinks=true, linkcolor=blue, citecolor=blue,
urlcolor=black}
\usepackage{url}
\usepackage{caption}
\arttype{Article}

\title{Intrusion-aware Alert Validation Algorithm for Cooperative Distributed Intrusion Detection Schemes of Wireless Sensor Networks}

\author{Riaz Ahmed Shaikh $^{1}$, Hassan Jameel$^{2}$, Brian J. d'Auriol$^{1}$, Heejo Lee$^{3}$, Sungyoung Lee$^{1,\star}$ and Young-Jae Song$^{1}$}

\address{\hangafter=1\hangindent=0.7em\noindent
$^{1}$ Department of Computer Engineering, Kyung Hee University, Suwon, Korea; E-Mails: riaz@oslab.khu.ac.kr (R.A.S.); daurial@oslab.khu.ac.kr (B.J.D.); yjsong@khu.ac.kr (Y.J.S.)\\
$^{2}$ Computing Department, Macquarie University, NSW, Australia; E-Mail: hasghar@science.mq.edu.au \\
$^{3}$ Department of Computer Science \& Engineering, Korea University, Seoul, Korea; E-Mail: heejo@korea.ac.kr \\[12pt]
$^{\star}$ Author to whom correspondence should be addressed;
E-Mail: sylee@oslab.khu.ac.kr; Tel.:+82-31-201-2514; Fax:
+82-31-202-2520
\\[12pt]
{\em Received: 24 April 2009; in revised form: 25 June 2009/
Accepted: 17 July 2009/ \\Published: 27 July 2009}}

\abstract{Existing anomaly and intrusion detection schemes of
wireless sensor networks have mainly focused on the detection of
intrusions. Once the intrusion is detected, an alerts or claims will
be generated. However, any \emph{unidentified} malicious nodes in
the network could send faulty anomaly and intrusion claims about the
legitimate nodes to the other nodes. Verifying the validity of such
claims is a critical and challenging issue that is not considered in
the existing cooperative-based distributed anomaly and intrusion
detection schemes of wireless sensor networks. In this paper, we
propose a validation algorithm that addresses this problem. This
algorithm utilizes the concept of intrusion-aware reliability that
helps to provide adequate reliability at a modest communication cost.
In this paper, we also provide a security resiliency analysis of the
proposed intrusion-aware alert validation algorithm.}

\keywords{Alerts; Anomalies; Intrusions; Trust Management; Wireless sensor networks}

\begin{document}



\section{Introduction}

Many anomaly and intrusion detection schemes (IDS) have been
proposed for wireless sensor networks~(WSNs)
~\cite{Bhuse06,WDu06,CELoo06,VChat07,Silva05,LiuF2007}, but those
schemes mainly focus on the detection of malicious or faulty nodes.
All those anomaly and intrusion detection schemes (IDS) that are
cooperative in nature ~\cite{Bhuse06,WDu06,VChat07} need to share
anomalies or intrusion claims with the other node(s). However, those
schemes are unable to ascertain that the alert or claim received by
the other node(s) is in fact sent by the trusted node(s). As a
result, any \textit{unidentified} malicious node(s) in the network
could send faulty anomaly and intrusion claims about the legitimate
node(s) to the other node(s). Verifying the validity of such claims
is a critical issue that is not considered in existing
cooperative-based distributed anomaly and IDS schemes of
WSNs~\cite{Bhuse07}. Recently, some intrusion prevention schemes
that are based on alerts have been proposed in the
literature~\cite{WTSu07,Zhang08}. However, these schemes are based
on the assumption that the monitoring nodes are trusted or the claim
will be trusted if the monitoring node passed simple authentication
and integrity test based on shared pair-wise key.

In this paper, we propose a new intrusion-aware alert validation
algorithm that provides a mechanism for verifying anomaly and
intrusion claims sent by any unidentified malicious node(s). This
algorithm is simple and easy to implement. Our proposed algorithm
execute on alert sender monitoring nodes and alert receiver
monitoring nodes. Sender monitoring nodes are mainly responsible for
the detection of malicious nodes, assignment of threat level, and
generation of alert messages, whereas receiver monitoring nodes are
mainly responsible for the validation of alert messages. Validation
mechanism consists of two phases: consensus phase and decision
phase. Although the consensus approach is widely used in distributed
computing domain to solve many problems like
fault-tolerance~\cite{Barborak93}, here we used this approach with
variation to solve problem of trusting anomaly and intrusion claims.
In consensus phase, we uniquely introduce an intrusion-aware
reliability concept that helps to provide an adequate reliability at
a modest communication cost. In the decision phase, a node will make
the decision regarding validation and invalidation of a claim based
on the result of consensus phase.

The rest of the paper is organized as follows: Section 2 contains
description on taxonomy of IDS. Section 3 describes related work.
Section 4 discusses the network model, assumptions and definitions.
Section 5 describes the proposed validation algorithm. Section 6
provides the analysis and evaluation of proposed algorithm  in terms
of communication overhead, reliability and security. Finally,
Section 7 concludes the paper and highlights some future work.

\section{Taxonomy of IDS}
\begin{figure}[!ht]
    \centering
    \caption{Taxonomy of intrusion detection schemes.}
     \label{fig:taxonomy}
        \includegraphics[width=3.8in]{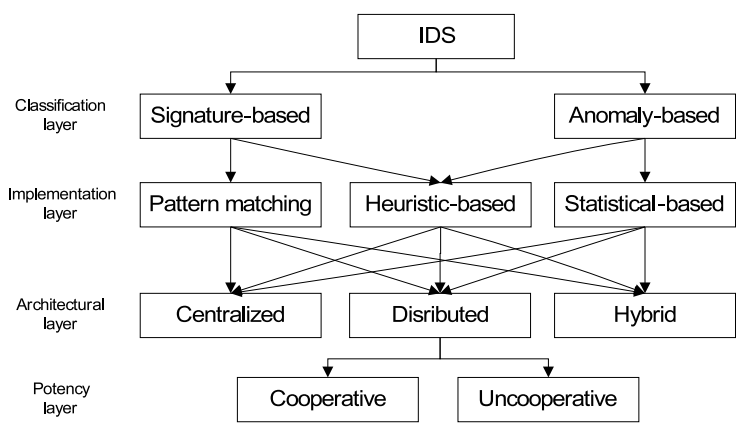}
\end{figure}

From the classification point of view, IDS have often been
categorized into two types: signature-based IDS and anomaly-based
IDS as shown in Figure~\ref{fig:taxonomy}. The signature-based IDS
schemes (mostly implemented via pattern matching approach) detect
intrusions based on the attack's signature, such as, specific byte
sequence in the payload or specific information in the header fields
like sender address, last hop address, etc. On the other hand, the
anomaly-based IDS (mostly implemented via statistical approach),
first determines the normal network activity and then checks all
traffic that deviates from the normal and marks it as anomalous.

\newpage In order to strengthen the signature-based and anomaly-based IDS
schemes, some researchers applied heuristic algorithms. Heuristic
approaches are generally used in AI. Instead of looking for exact
pattern matches or simple thresholds, heuristic-based IDS ``looks
for behavior that is out of ordinary"~\cite{Shari2003} during
specific time interval. In simple words, it ``uses an algorithm to
determine whether an alarm should be fired''~\cite{Newman2004}. For
example, if a threshold number of unique ports are scanned on a
particular host or a specific attack pattern signature is detected,
then alarm will be fired~\cite{Newman2004}.

From an architectural point of view, IDS schemes are further
categorized into three categories: centralized, distributed and
hybrid. In the centralized approach, a single designated node
monitors the whole network. In the distributed approach, every node
or a group of nodes monitor the network. In the hybrid approach,
every group has one selected primary node responsible for monitoring
and detecting anomalies and intrusions. Once the information is
gathered, it is forwarded to the central base station which
calculates the impact of those anomalies and intrusions on the whole
network.

From the potency point of view, distributed approach is further
classified into cooperative and uncooperative distributed
approaches. In the cooperative distributed approach, every node or a
group of nodes exchanges information about the anomalies and
intrusions in order to detect collaborative intrusion attacks. On
the contrary, in the uncooperative distributed approach, nodes do
not share information about anomalies and intrusion with each
others.

\section{Related Work}

\subsection{Intrusion Detection Schemes}
Intrusion detection schemes are not in itself the main focus of this
paper. However, in order to give a brief overview of those, we have
summarized the existing proposed anomalies and IDS schemes of WSNs
in Table~1, in which~\cite{Bhuse06},~\cite{WDu06},~\cite{VChat07}
and ~\cite{LiuF2007} are distributed and cooperative in nature.
Brief descriptions of some of the proposed schemes are given below.

Bhuse et al.~\cite{Bhuse06} have proposed different lightweight
techniques for detecting anomalies for various layers, such as
application, network, MAC and physical. The main advantage of the
proposed techniques is the low overhead that makes them energy
efficient. This is due to the fact that they reuse the already
available system information (e.g. RSSI values, round trip time,
etc.) which are brought forth at various layers of network stack.

\begin{table} [ht!]
\renewcommand{\arraystretch}{1.5}
    \caption{Summarization of proposed Anomalies and IDS schemes of WSNs}
    \label{tab:mem}
    \centering
\scalebox{0.7}{
   \begin{tabular}{|c|c|c|c|c|c|c|c|}
           \hline
    \multicolumn{2}{|c|}{} & \cite{Bhuse06} & \cite{WDu06} & \cite{CELoo06} & \cite{VChat07}  & \cite{Silva05} & \cite{LiuF2007}  \\ \hline
   &Technique&\parbox{1.1in}{Signature-based}&\parbox{1.1in}{Statistical-based}&\parbox{1.1in}{Statistical-based}&\parbox{1.1in}{Statistical-based}&\parbox{1.1in}{Statistical-based} & \parbox{1.125in}{Statistical-based} \\ \cline{2-8}

  Classification     &Architecture&\parbox{1.1in}{Distributed \& cooperative}&\parbox{1.1in}{Distributed \& cooperative}&\parbox{1.1in}{Distributed \& uncooperative}& Hybrid&\vspace{5pt}\parbox{1.1in}{Distributed \& uncooperative} & \parbox{1.1in}{Distributed \& cooperative} \\ \hline

     &\parbox{1in}{Installation of IDS}&\parbox{1.1in}{Each sensor node}&\parbox{1.1in}{Each sensor node}&\parbox{1.1in}{Each sensor node}&\parbox{1.1in}{Each primary node of a group}&\vspace{5pt}\parbox{1.1in}{Special monitor nodes in network} &\parbox{1.1in}{Each sensor node} \\ \cline{2-8}

Specifications  & IDS Scope&\parbox{1.1in}{Multilayer (Appl., Net.,
MAC \& Phy.)}&\vspace{5pt}\parbox{1.1in}{Application
layer}&\parbox{1.1in}{Network layer}&\parbox{1.1in}{Application
layer}&\parbox{1.1in}{Multilayer (Appl., Net., MAC \& Phy.)}
&\parbox{1.1in}{Network layer} \\ \cline{2-8}

     &\parbox{1in}{Attacks detects}&\vspace{5pt}\parbox{1.1in}{Masquerade attack, and forged packets attacks}&\parbox{1.1in}{Localization anomalies}&\parbox{1.1in}{Routing attacks e.g. Periodic error route attack, active \& passive sinkhole attack}&\parbox{1.1in}{Correlated anomalies / attacks (invalid  data insertion)}&\parbox{1.1in}{Worm hole, data alteration, selective forwarding, black hole, \& jamming} & \parbox{1.1in}{Routing attacks e.g. packet dropping etc.}\\ \hline

Network &\parbox{1in}{Sensor node}&Static / Mobile& Static & Static
/ Mobile& Static / Mobile& Static & Static\\ \cline{2-8}
     &Topology& Any& Any& Any& Cluster-based& Tree-based & Any\\ \hline
        \end{tabular}}
\end{table}

Chatzigiannakis et al.~\cite{VChat07} have proposed an application
level anomaly detection approach that fuses data (comprised of
multiple metrics) gathered from different sensor nodes. In the
proposed scheme, the authors have applied Principal Component
Analysis (PCA) to reduce the dimensionality of a data set. So this
approach will help to detect correlated anomalies/attacks that
involve multiple groups of sensors.

Du et al.~\cite{WDu06} have proposed a localization anomalies
detection (LAD) scheme for the wireless sensor networks. This scheme
takes the advantage of the deployment knowledge and the group
membership of its neighbors, available in many sensor network
applications. This information is then utilized to find out whether
the estimated location is consistent with its observations. In case
of an inconsistency LAD would report an anomaly.

Loo et al.~\cite{CELoo06} have proposed an anomaly based intrusion
detection scheme that is used to detect network level intrusions,
e.g., routing attacks. They use clustering algorithm to build the
model of normal network behavior, and then use this model to detect
anomalies in traffic patterns. IDS will be installed on each sensor
and each IDS will function independently.

Da Silva et al.~\cite{Silva05} have proposed high level methodology
to construct the decentralized IDS for wireless sensor networks.
They have adopted statistical approach based on the inference of the
network behavior. The network behavior is obtained from the analysis
of the events detected at the specific monitor node, which is
responsible for monitoring its one-hop neighbors looking for
intruder(s).

Liu et al.~\cite{LiuF2007} have proposed insider attack detection
scheme for wireless sensor networks. They have adopted localized
distributed and cooperative approach. This scheme explores the
spatial correlation in neighborhood activities and requires no prior
knowledge about normal or malicious nodes. This scheme works in four
phases: (1) collection of local information about neighborhood nodes
(e.g., packet dropping rate, sending rate, etc.), (2) filtering the
collected data, (3) identification of initial outlying (malicious)
nodes, and (4) applying majority vote to obtain a final list of
malicious nodes. Once the node detects some malicious node, it will
forward the report to the base station. Afterwards the base station
will isolate that node from the network.

\subsection{Intrusion Prevention Schemes}

Su et al.~\cite{WTSu07} have proposed an energy-efficient Hybrid
Intrusion Prohibition (eHIP) system for cluster-based wireless
sensor networks. The eHIP system consists of two subsystems:
Authentication-based Intrusion Prevention (AIP) subsystem and
Collaboration-based Intrusion Detection (CID) subsystem. In AIP, two
distinguish authentication mechanisms are proposed to verify the
control and sensed data messages with the help of HMAC and the
modified form of one-key chain~\cite{Perrig2000} mechanisms. CID is
also consisted of two subsystems: cluster head monitoring (CHM)
system and member node monitoring (MNM) system. In CHM, all member
nodes are divided into multiple monitoring groups. With respect to
security requirements, each monitoring group has various number of
monitoring nodes. Every monitoring group monitors the cluster head.
Whenever any monitoring group detects malicious activity of the
cluster head, it generates an alarm that is forwarded to all member
nodes of the cluster. Each member node maintains the alarm table. If
the number of alarms exceeds then the alarm threshold, the cluster
head will be declared as a malicious node. The member node
monitoring mechanism is performed at the cluster head and limited to
the detection of compromised nodes through the used pair-wise key
only.

Zhang et al.~\cite{Zhang08} have proposed a nice
application-independent framework for identifying compromised nodes.
This framework is based on alerts generated by specific intrusion
detection system. The authors have adopted a centralized approach
and used a simple graph theory. However, this scheme has some
limitations, e.g., it provides some late detection of compromised
nodes, because the detection process will always start at the end of
each time window. If the size of the time window is large (e.g., one
hour, as mentioned in~\cite{Zhang08}), then it is very likely that
an adversary can achieve its objective during that time window. If
the time window is small, then the result may not be accurate. Also,
the detection accuracy is mainly dependent on the size of the
network density. If the network size decreases, then the detection
accuracy will also decrease.

\section{Network Model, Assumptions and Definitions}

\subsection{Network Model and Assumptions}

Sensor nodes are deployed in an environment either in a random
fashion or in a grid fashion. After deployment nodes become static,
nodes are organized into clusters. The reason behind taking
cluster-based network model is that it is widely used in real world
scenarios for efficient network organization~\cite{Younis06}. Within
a cluster, communication mechanism could be
single-hop~\cite{Heinzelman02} or multi-hop~\cite{YanJin08}. In case
of a multi-hop clustering environment, the cluster could be divided
into two or three sensor sub-clusters for the purpose of distributed
detection~\cite{AnnaHac2003}.

We assume that any cooperative-based distributed anomaly detection
or IDS is already deployed in the WSNs and forwards claims to the
other node(s) whenever it detects anomalies or intrusions. Based on
the mechanism of the IDS, every node or subset of nodes (within a
cluster) acts as a monitoring node. The malicious node must fall
into the radio range of the monitoring node. And the node (who
received the claim from the monitoring node) has the knowledge about
the neighboring nodes of the monitoring and malicious nodes (the
malicious and monitoring nodes belong to the same cluster). Within a
cluster or sub-cluster, all monitoring nodes can communicate with
each other directly. We also assume that the multiple sensor nodes
in a neighborhood can sense the same anomaly/intrusion. We also
assume that all information is exchanged in a secure encrypted
manner. For this purpose, every monitoring node share a unique
secret key~\cite{Riaz2006} with other monitoring node(s) that are in
the same cluster.

\subsection {Definitions}

A legitimate node compromised by an adversary is called a malicious
node. In order to hide the presence of the adversary, a malicious
node could also perform all activities of the normal nodes, such as
monitoring, ciphering of data, forwarding of packets, etc.

\begin{figure}[h]
    \centering
    \caption{Intrusion-aware reliability mode concept.}
    \label{fig:IAR}
        \includegraphics[width=3.5in]{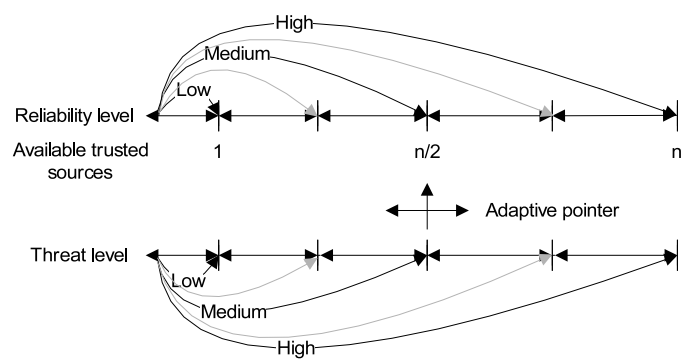}
    \end{figure}

Reliability means the confidence level on a certain decision. It can
simply be categorized into three levels: (1) low, (2) medium, and
(3) high. At low reliability, validation is based on the
confirmation from any single available trusted source. At medium
reliability, validation is based on the confirmation from half of
the available trusted sources. At high reliability, validation is
based on the confirmation from all of the trusted sources. In
general, the reliability level ($R_L$) is define as:
\begin{equation}
R_L  = q\,\,\,;\,\,\,\,\,q \leq n
\end{equation}
where $n$ represents the total number of available trusted nodes,
and $q$ represents the number of nodes consulted. However, in order
to achieve more flexibility and adaptability, we have adopted the
intrusion-aware reliability mode concept, in which the validation is
based on the level of a threat of an anomaly or intrusion. This
approach will also reduce the communication cost as described in
Section 6.1. Threats could also be categorized into low, medium,
high or other. Depending on the level of the threat, intrusion-aware
reliability mode is set to low, medium, high or other, as shown in
Figure~\ref{fig:IAR}.

\section{Intrusion-aware Alert Validation Algorithm}

Our proposed intrusion-aware alert validation algorithm execute on
sender as well as on receiver monitoring nodes. Sender monitoring
nodes are mainly responsible for the detection of malicious nodes
and generation of alert messages, whereas receiver monitoring nodes
are mainly responsible for the validation of alert messages. Both
sender and receiver nodes go through different phases as \linebreak
described below.

\subsection{Sender Monitoring Node}

\begin{figure}[h]
    \centering
     \caption{Intrusion-aware validation algorithm at sender end.}
    \label{fig:AlgorithmPhase1}
        \includegraphics[width=0.35\textwidth]{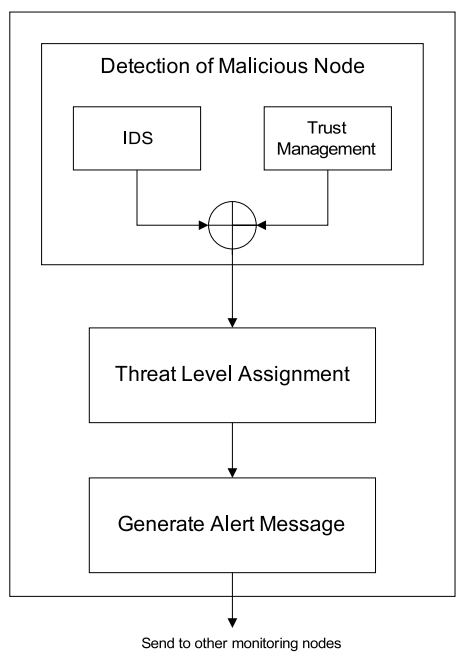}
\end{figure}

In our proposed algorithm, sender monitoring node will perform
mainly three steps (as shown in Figure~\ref{fig:AlgorithmPhase1}):
(1) detection of malicious node, (2) threat level assignment, and
(3) generation of alert message.

\subsubsection{\textbf{Phase 1: Detection of Malicious Node}}

A node can be classified into one of the three
categories~\cite{Riaz2006b}: trustworthy, untrustworthy, and
uncertain. A node is considered to be trustworthy if it interacts
successfully most of the time with the other nodes. A node is
considered untrustworthy if it tries to do as many unsuccessful
interactions as possible with the other nodes. An untrustworthy node
could be a faulty~\cite{Jiang09} or malicious node. A node is
considered uncertain if it performs both successful and unsuccessful
interactions. Detailed definition of the successful and unsuccessful
interactions and trust calculation methodology is available in our
paper~\cite{Riaz2009} and provided here in a simplified form.

A sender will consider an interaction successful if the sender
receives confirmation that the packet is successfully received by
the neighbor node and forwarded towards the destination in an
unaltered fashion. The first requirement of successful reception is
achieved on the reception of the link layer acknowledgment (ACK).
The second requirement of forwarding towards the destination is
achieved with the help of enhanced passive acknowledgment (PACK) by
overhearing the transmission of a next hop on the route, since they
are within the radio range~\cite{Buchegger04}. If the sender node
does not overhear the retransmission of the packet within a
threshold time from its neighboring node or the overheard packet is
found to be illegally fabricated (by comparing the payload that is
attached to the packet), then the sender node will consider that
interaction as unsuccessful.

With the help of this simple approach, several attacks can be
prevented, i.e., the black hole attack is straightforwardly detected
when malicious node drops the incoming packets and keeps sending
self-generated packets~\cite{Gupta06}. Similarly, sink hole
attack~\cite{DuX07}, an advanced version of the black hole attack,
is also easily detectable by looking at the passive acknowledgment.
Likewise, the selective forwarding attack~\cite{Karlof03} and
gray-hole attack~\cite{srinivasan06} can also be eliminated with the
aid of above mentioned approach.

Based on these successful and unsuccessful interactions, node $x$
can calculate the trust value of node $y$:
\begin{equation}
 T_{x,y}  = \left[ {100\left( {\frac{{S_{x,y} }}{{S_{x,y}  + U_{x,y} }}} \right)\left( {1 - \frac{1}{{S_{x,y} + 1}}} \right)} \right]
\end{equation}
where [.] is the nearest integer function,  $S_{x,y}$ is the total
number of successful interactions of  node $x$ with $y$ during time
$\Delta t$, $U_{x,y}$ is the total number of unsuccessful
interactions of node $x$ with $y$ during time $\Delta t$. After
calculating trust value, a node will quantize trust into three
states as follows:
\begin{equation}
\label{eq:bound}
\small Mp(T_{x,y} ) = \left\{ {\begin{array}{*{20}c}
   \textnormal{trustworthy} & {100- f \le T_{x,y} \le 100}  \\
   \textnormal{uncertain} & {50- g \le T_{x,y} < 100 - f}  \\
   \textnormal{untrustworthy} & {0 \le T_{x,y} < 50 - g}  \\
\end{array}} \right\}
\end{equation}
where, $f$ represents half of the average values of all trustworthy
nodes and $g$ represents one-third of the average values of all
untrustworthy nodes. Both $f$ and $g$ are calculated as follows:
\begin{equation}
f_{j+1} = \left\{ {\begin{array}{*{20}c}
   \left[{\frac{1}{2}\left( {\frac{{\sum _{i \in R_x } T_{x,i} }}{{\left| {R_x } \right|}}} \right)}\right] & {0 < \left| {R_x } \right| \le n - 1}  \\
   {f_j} & {\left| {R_x } \right| = 0}  \\
\end{array}} \right.
    \label{Eq:f}
\end{equation}
\begin{equation}
g_{j+1} = \left\{ {\begin{array}{*{20}c}
   \left[{\frac{1}{3}\left( {\frac{{\sum _{i \in M_x } T_{x,i} }}{{\left| {M_x } \right|}}} \right)}\right] & {0 < \left| {M_x } \right| \le n - 1}  \\
   {g_j} & {\left| {M_x } \right| = 0}  \\
\end{array}} \right.
    \label{Eq:g}
\end{equation}
where [.] is the nearest integer function, $R_x$ represents the set
of trustworthy nodes for node $x$, $M_x$ the set of untrustworthy
nodes for node $x$, and $n$ is the total number of nodes that
contains trustworthy, untrustworthy and uncertain nodes. The initial
trust values of all nodes are 50, which represents the uncertain
state. Initially $f$ and $g$ are equal to 25 and 17 respectively,
although other values could also be used by keeping the following
constraint intact: $f_i - g_i \geq 1 $, which is necessary for
keeping the uncertain zone between a trusted and untrustworthy zone.
The values of $f$ and $g$ are adaptive. During the steady-state
operation, these values can change with every passing unit of time
which creates dynamic trust boundaries. At any stage, when $|R_x|$
or $|M_x|$ becomes zero, the value of $f_{j+1}$ or $g_{j+1}$ remains
the same as the previous values ($f_j$ and $g_j$). The nodes whose
values are above $100-f$ will be declared as trustworthy nodes
(Equation ~\ref{eq:bound}), and nodes whose values are lower than
$50-g$ will be consider as untrustworthy nodes (Equation
~\ref{eq:bound}). After each passage of time, $\Delta t$, nodes will
recalculate the values of $f$ and $g$. This trust calculation
procedure will continue in this fashion.

The time window length ($\Delta t$) could be made shorter  or longer
based on the network analysis scenarios. If $\Delta t$ is too short,
then the calculated trust value may not reflect the reliable
behavior. On the other hand, if it is too long, then it will consume
too much memory to store the interaction record at the sensor node. 
Therefore, various parameters can be used to adjust the length of
$\Delta t$. In simplicity, let us assume a cluster size, $n$, as the 
single parameters; then, $\Delta t$ is equal to $n-1$. This approach 
reduces the problems associated with too short or too long time window 
lengths. Moreover, the time window lengths are adaptive based on the 
cluster size. If the size of the cluster changes, then the time window length will also change.

\subsubsection{\textbf{Phase 2: Threat level Assignment}}

Whenever a monitoring node detects malicious node, it generates an
alert. This alert will be forwarded to the other neighbor monitoring
nodes. In our proposed algorithm, an alert message also contains the
information about intensity or level of threat. This level of threat
is calculated based on the trust value of the malicious node. The
lesser the trust value, the higher the threat level. For example, if
a node is continuously dropping all incoming packets (black hole
attack), then based on the trust management methodology defined
above, the trust value of a malicious node becomes zero. So, the
level of threat for this kind of attack is high. If a node is
performing sink hole attack, then the trust value of a node become
higher than the node performing the black hole attack. Therefore,
the threat level is less as compared with the earlier one.

Based on the trust value of a malicious node, a node will quantize threat level ($H_{level}$) in following way:
\begin{equation}
H_{level}  = \left\{ {\begin{array}{*{20}c}
   {H_1 } & {(k - 1) \times \frac{{50 - g}}{k} \le T_{mal}  \le 50 - g}  \\
   {H_2 } & {(k - 2) \times \frac{{50 - g}}{k} \le T_{mal}  < (k - 1) \times \frac{{50 - g}}{k}}  \\
    \vdots  &  \vdots   \\
   {H_{k - 1} } & {\frac{{50 - g}}{k} \le T_{mal}  < 2 \times \frac{{50 - g}}{k}}  \\
   {H_k } & {0 \le T_{mal}  < \frac{{50 - g}}{k}}  \\
\end{array}} \right\}
\end{equation}
where $k$ represents total number of threat levels, $50-g$ represent
the upper limit of the untrustworthy zone as defined in
Equation~\ref{eq:bound}. $T_{mal}$ represent the trust value of a
malicious node. Let us assume that there are three threat levels
($k$=3): low, medium and high. In that case, a node will quantize
threat level ($H_{level})$ in following way:
\begin{equation}
H_{level}  = \left\{ {\begin{array}{*{20}c}
   {{\rm{Low}}} & {2 \times \frac{{50 - g}}{3} \le T_{mal}  \le 50 - g}  \\
   {{\rm{Medium}}} & {\frac{{50 - g}}{3} \le T_{mal}  < 2 \times \frac{{50 - g}}{3}}  \\
   {{\rm{High}}} & {0 \le T_{mal}  < \frac{{50 - g}}{3}}  \\
\end{array}} \right\}
\end{equation}
The concept of threat level is later used in our algorithm for a selection of an appropriate reliability level.

\subsubsection{\textbf{Phase 3: Generation of Alert Message}}

Once the threat level is assigned, a node will generate an
alert/claim message. This message contains four types of
information.
\begin{enumerate}
    \item Identity of the sender node ($ID_{sender}$).
    \item Identity of the malicious node ($ID_{mal}$).
    \item Threat level ($H_{level}$).
    \item Threat detail, like code etc.
\end{enumerate}
This message will be forwarded to the other monitoring nodes.

\subsection{Receiver Monitoring Node}

\begin{figure}[h]
    \centering
        \caption{Intrusion-aware validation algorithm at the receiver end.}
    \label{fig:algorithm}
        \includegraphics[width=0.75\textwidth]{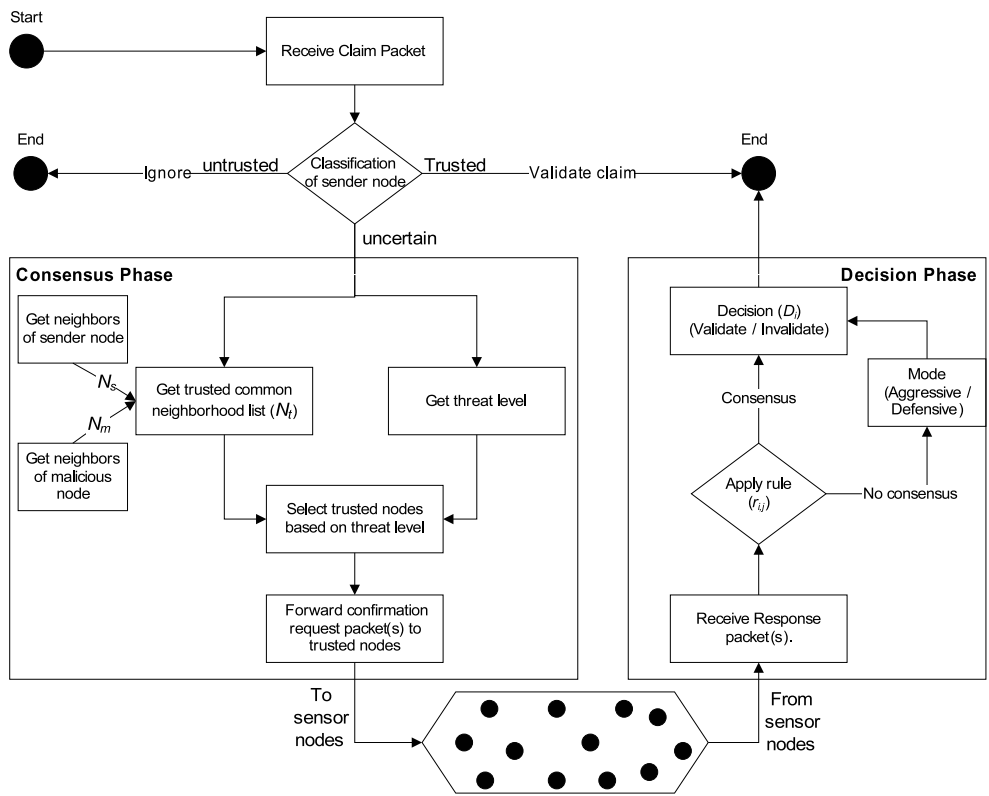}
\end{figure}

Intrusion-aware alert validation algorithm at the receiver end is
shown in Figure~\ref{fig:algorithm}. It shows that, if the claim
packet is received from \textit{trustworthy} monitoring node, then
claim will be validated straightforwardly. If the claim packet is
received from the \textit{untrustworthy} monitoring node, then no
consideration will be given to that claim packet. If the claim
packet is received from the \textit{uncertain} monitoring node, then
our proposed intrusion-aware validation algorithm goes through two
phases: (1) consensus phase, and (2) decision phase. Details about
these phases are given below.

\subsubsection{Phase 1 (Consensus Phase)} 
Whenever a designated node receives the claim/alert packet, it first
checks (1) if the sender is uncertain, and (2) if the identity of a
new malicious node is already declared as a malicious node
(Algorithm 1, Line 1:2). If not, then the node will first get the
common neighborhood list ($N_{sm}$) of  the sender and malicious
nodes respectively~(Line 3:5). Afterwards, the node will perform
filtering by eliminating any known malicious node(s) from that
list~(Line~6). Based on the threat level, confirmation request
packet(s) is forwarded to randomly selected node(s) from the $N_t$
list~(Line~7:19). For example, if the threat level is low, then the
confirmation request is forwarded to the one randomly selected
trustworthy node from the list $N_t$ (Line 8:9). If the threat level
is medium, then the confirmation request packet is forwarded to half
of the randomly selected trustworthy nodes from the list $N_t$ (Line
10:13). If the threat level is high, then the confirmation request
packet is forwarded to all trustworthy nodes in the list $N_t$ (Line
14:17). If the information about the malicious node is already
present~(Line 20), then the node will just update its old record
(Line 21).

\begin{algorithm}[ht]
\label{ValALGO}
\caption{Phase 1: Consensus Phase}
\begin{algorithmic}[1]
    \STATE Received Claim Packet ($ID_{sender},ID_{mal}, H_{level}$,detail);
    \IF { $ID_{sender}$ is uncertain and $ID_{mal}$ is new}
        \STATE $N_s$ = GetNeighorList($ID_{sender}$);
        \STATE $N_m$ = GetNeighorList($ID_{mal}$);
        \STATE $N_{sm}$ = $N_s \bigcap N_m$;
        \STATE $N_t$ = Eliminate\_Known\_Malicious\_Nodes($N_{sm}$);
        \IF { $N_t \neq \phi $}
          \IF { ThreatLevel($TH_{level}$) is Low }
            \STATE Send conf\_req\_pkt($rand(N_t),ID_{mal},H_{level}$,det);
          \ELSIF {ThreatLevel($TH_{level}$) is Medium}
            \FOR{ $i=1 \,\, to \,\, len(N_t)/2 $}
              \STATE Send conf\_req\_pkt($rand(N_t),ID_{mal},H_{level}$,det);
            \ENDFOR
          \ELSE
            \FOR{ $i=1 \,\, to \,\, len(N_t) $}
               \STATE Send conf\_req\_pkt($ID_i,ID_{mal}$,det);
            \ENDFOR
          \ENDIF
        \ENDIF
  \ELSE
     \STATE Update Record;
    \ENDIF
    \end{algorithmic}
\end{algorithm}

\subsubsection{Phase 2 (Decision Phase)} Once the confirmation request packet(s) is forwarded to the particular node(s) then the phase 2 of the validation algorithm is triggered. In this phase algorithm will first wait for the confirmation response packets until $\delta t$ time, where $\delta t$ is calculated as:
\begin{equation}
\delta t = 2 \left[ 2t_{prop} + t_{proc} \right]
\end{equation}
Here, $t_{prop}$ is the propagation time between the requester and
farthest responder (in terms of hops or geographical location) among
nodes where the request packets were forwarded. The $t_{proc}$ is
the estimated processing time of the request at the responder end.

A node will expect three types of responses ($r$) from the nodes where confirmation request packets were forwarded:
\begin{equation}
r_{i,j}  = \left\{ {\begin{array}{*{20}c}
   1  & \rm{if} & \rm{agree\,with\, claim}  \\
   0  & \rm{if} & \rm{don't\,know}  \\
   -1 & \rm{if} & \rm{not\,agree\,with\,claim}  \\
\end{array}} \right.
\end{equation}
where $r_{i,j}$ represents that the node $i$ received the response
packet from the node $j$ and $j \in N_t$. A node $i$ will make the
decision ($D$) about the validity and invalidity of the claim based
on the following rule:
\begin{equation}
D_i  = \left\{ {\begin{array}{*{20}c}
   \rm{validate} & \rm{iff} & {\sum\limits_{j = 0}^{n_{res} } {r_{i,j} }  > 0}  \\
   \rm{no\,consensus} & \rm{iff} & {\sum\limits_{j = 0}^{n_{res} } {r_{i,j} }  = 0}  \\
   \rm{invalidate} & \rm{iff} & {\sum\limits_{j = 0}^{n_{res} } {r_{i,j} }  < 0}  \\
\end{array}} \right.
\end{equation}
where $n_{res}$ represents the total number of the response packets
received by the node $i$ in response to the number of the request
packets ($n_{req}$). Here $0 \ge n_{res} \leq n_{req}$.

If the claim is invalidated, then the sender of the claim will
declare it as a malicious node. That helps to provide protection
against any possible security threats, such as flooding, denial of
service attacks, etc.

If no consensus is available, then the algorithm will decide based
on its mode that is set by the administrator. There are two types of
modes: aggressive and defensive. If the algorithm is set in the
aggressive mode, then the node will validate the claim; if it is set
in the defensive mode, then the node will invalidate the claim.

\textbf{Responder monitoring nodes}: Whenever any monitoring node
receives confirmation request packet for alert validation, it will
first check the status of the sender. If the sender is trusted, it
will generate confirmation response packet and will not generate the
same alert if responder node agree with the claim. Also, the
responder node will update its malicious node list. If the responder
node receives the same alert message from another monitoring node,
it will straightforwardly validate that claim. This approach helps
to suppress any extra requests for the same alert.

\textbf{Tolerance for false alarm}: In our proposed algorithm,
default tolerance level for false alarms generated by any node is
zero. As mentioned earlier, if a claim is invalidated, the sender of
the claim will be declared as a malicious node. If we do not declare
the sender as a malicious node, then it may result in flooding or
denial of service attacks. However, if we declare the sender as a
malicious node, it may cause a node to be evicted from the network
due to false alarm.

In order to solve the above problem, we can introduce a tolerance
level metric in our algorithm. Tolerance level determines the amount
of traffic each node can generate for the claims about which it is
unsure. Tolerance level will depend on network capacity and node
abundance. It may also depend on the energy level of the network. If
the energy level is too low, the application can decide not to
tolerate any such traffic.

\section{Analyses and Evaluation}
\subsection{Communication Overhead Analysis}
The communication overhead of the validation algorithm is dependent
on three factors: (1) total number of intrusion claims ($I_c$), (2)
number of commonly trusted neighboring nodes, and (3) threat level
of intrusion or anomaly. Table~\ref{tab:commoverhead} shows the
communication overhead, in which $m_t$ represents the average number
of trusted common neighboring nodes between the monitoring and
malicious nodes and $I_l, I_m$, and $I_h$ represents the total
number of low, medium and high intrusion level threats respectively.
Here $I_c = I_l + I_m + I_h$.
\begin{table}[h]
    \caption{Communication overhead of reliability modes.}
    \label{tab:commoverhead}
    \centering
    \begin{tabular}{|c|c|}
        \hline
            & Cost \\ \hline \hline
         Low &  $2I_c $ \\ \hline
         Medium & $m_tI_c$ \\ \hline
         High  & $2m_tI_c$  \\ \hline
         Intrusion-aware & $2I_l + (I_m + 2I_h)m_t$ \\ \hline
        \end{tabular}
\end{table}

\begin{figure}[!ht]
   \caption{Average communication overhead of validation algorithm after 1000 simulation runs in which different levels of intrusions occurs randomly.}
 \subfigure[Effect of $m_t$ and $I_c$]{
           \label{fig:fixedMN}
           \includegraphics[width=3.25in]{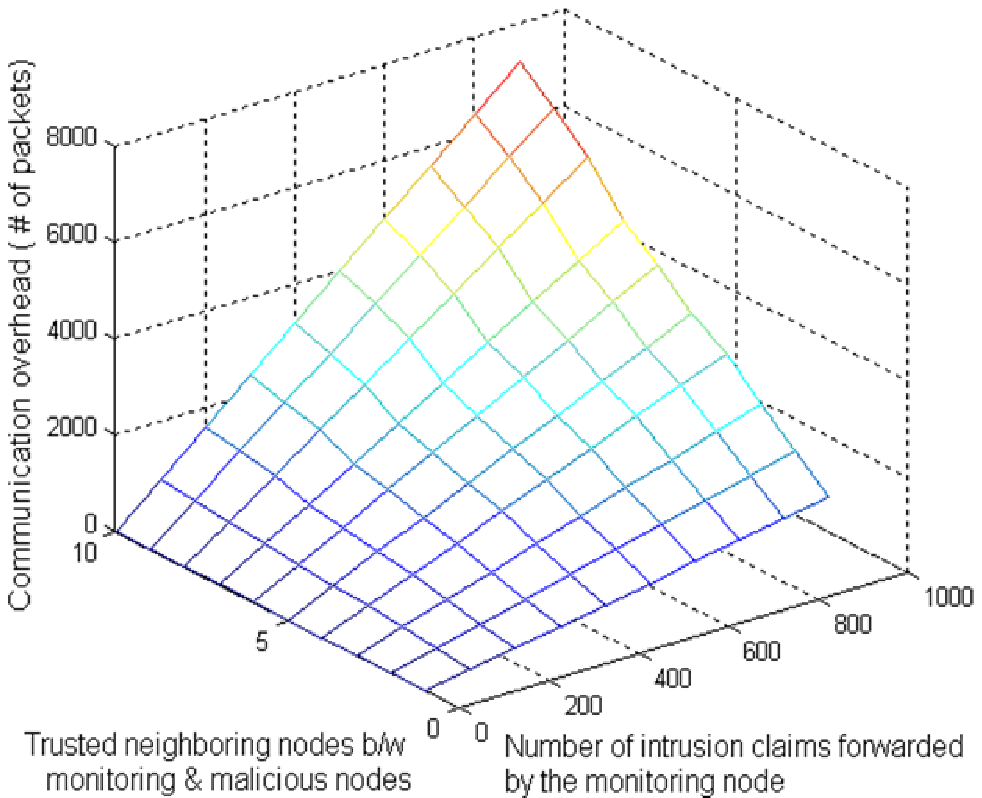}}
 \subfigure[Comparison from reliability mode perspective]{
           \label{fig:randomMN}
           \includegraphics [width=3.25in]{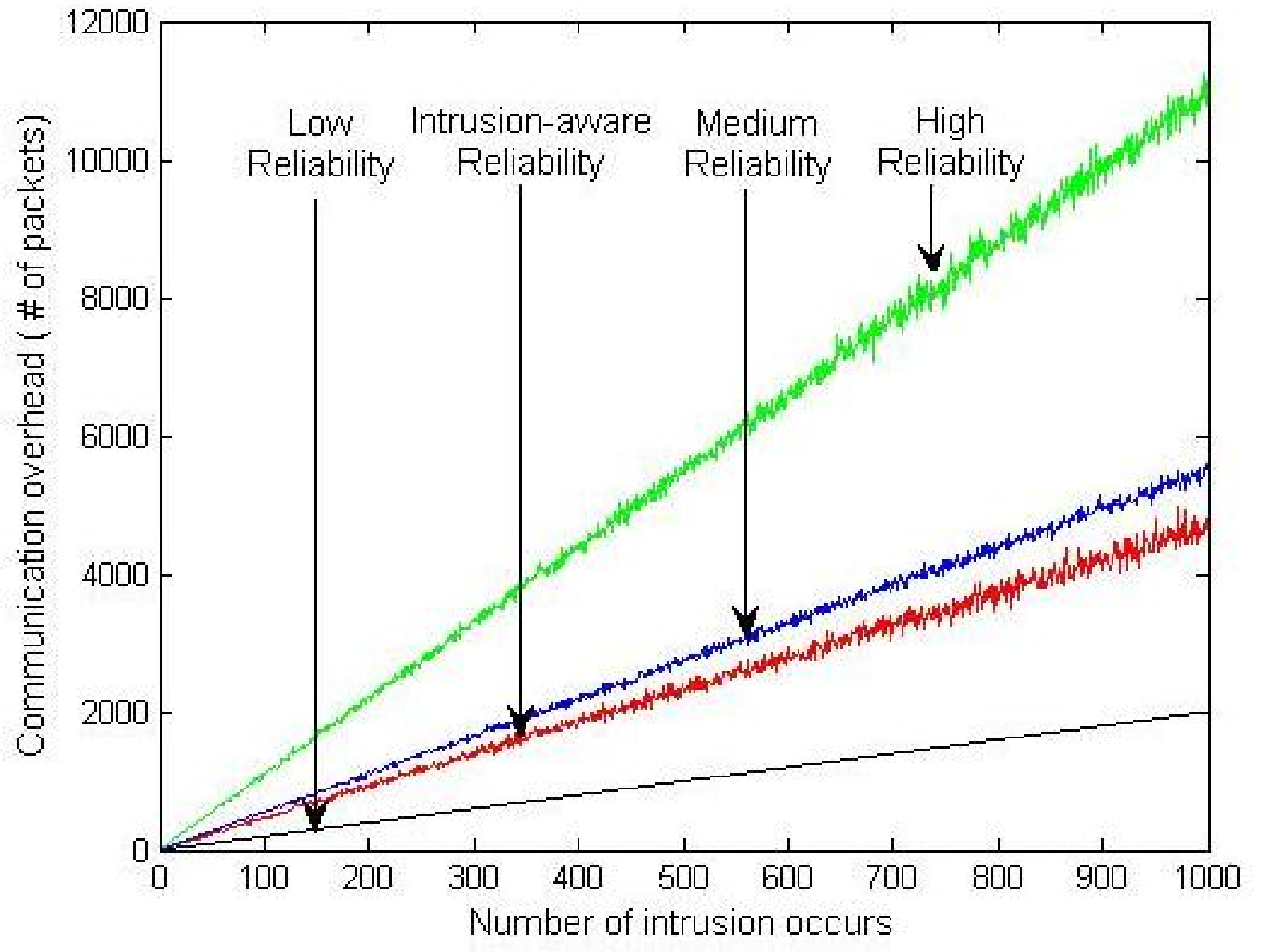}}
\label{fig:CO}
\end{figure}

Figure~\ref{fig:CO} shows the average communication overhead (1000
simulation runs) of the proposed validation algorithm. During the
simulation, different levels (low, medium or high) of threats of
anomalies and intrusions occur randomly. Figure~\ref{fig:fixedMN}
shows the effect of average number of commonly trusted neighboring
nodes (between the monitoring and the malicious nodes) $m_t$ and the
total number of intrusions $I_c$ occurred in the network. It shows
that as the number of $m_t$ or $I_c$ increases, the communication
overhead of the validation scheme also increases linearly.
Figure~\ref{fig:randomMN} shows the comparison between the four
different levels of the reliability modes. In the simulation, each
monitoring node has a random number of commonly trusted neighboring
nodes. This figure shows that the intrusion-aware reliability mode
introduces less communication overhead then the medium and high
level reliability modes. At a modest communication cost, it provides
adequate reliability required by the nature of the intrusion claim.

\begin{figure}[!ht]
    \centering
        \caption{Effect of false alarm tolerance factor on communication.}
    \label{fig:tolCommOverh}
        \includegraphics[width=3.75in]{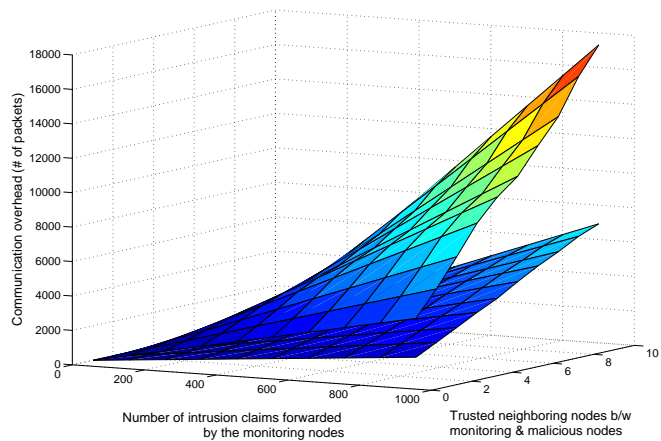}
\end{figure}

Figure~\ref{fig:tolCommOverh} shows the effect of tolerance level
for false alarms on communication overhead. As we mentioned earlier,
the default tolerance level in our proposed scheme is zero. During
simulation, we introduce four tolerance levels (0--3) that occurred
randomly. Figure~\ref{fig:tolCommOverh} shows that as the number of
$m_t$ or $I_c$ increases, the communication overhead of the
validation scheme (with random tolerance) increases more sharply as
compared with the zero tolerance level.

\subsection{Reliability Analysis}
If we assume that the responding node have equal probability of
sending any one of the three possible responses (agree, disagree and
don't know), then the total probability ($P_c$) of a algorithm to
reach at the consensus state (validate or invalidate) is:
\equation
P_c = \frac{{N_c }}{{K^{n_{res}}}}
\endequation
where $N_c$ represents the number of nodes reaching a consensus and
$K$ represents the number of possible outcomes (agree, disagree and
don't know) produces by the node. If the probability distribution is
not uniform between possible outcomes, then the total probability
($P_c$) of an algorithm to reach the consensus state (validates or
invalidate) is:
\begin{equation}
\begin{array}{l}
 P_c  = \sum\limits_{m = 1}^M {(\Pi _{i = 1}^{n_{res}} PMF_i (S_m (i))) \times \delta (m)} \\
 \mbox{ where } \,\, M = K^{n_{res}}
 \end{array}
\end{equation}
where $\delta (m)$ is one if $m$ node reaches the consensus, zero if
otherwise. $PMF_i$ is the probability mass function that captures
the probability distribution of the symbol produced by the node $i$.
$S_m (i)$ is the $i^{th}$ symbol in the $m^{th}$ node result. More
details and derivation of these two probability equations are given
in~\cite{Yacoub02}.

\begin{figure}[h]
    \centering
       \caption{Probability of reaching at consensus and no consensus state.}
    \label{fig:reliability}
        \includegraphics[width=3.35in]{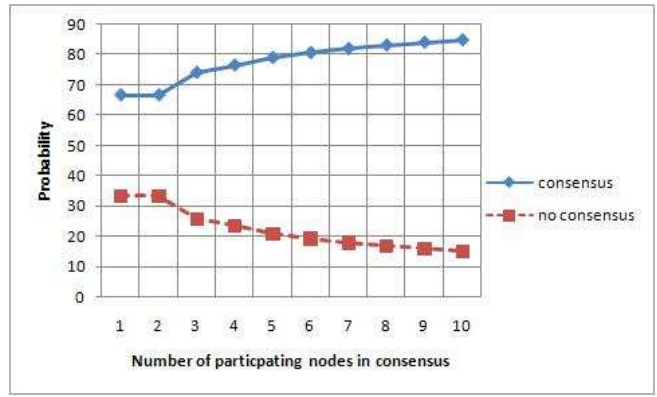}
\end{figure}

Figure~\ref{fig:reliability} shows the simulation result for the
probability of reaching consensus (validate or invalidate) of our
validation algorithm. It shows that as the number of participating
nodes increases in the consensus process, the probability of
reaching some consensus also increases linearly.

\subsection{Security Resiliency Analysis}

Let $\bf{s}$ denote the claimant node (the sender), let $\bf{m}$
denote the node accused of being malicious and let $\bf{a}$ denote
the node that receives this claim from $\bf{s}$ about $\bf{m}$. As
before, let $N_t$ denote the filtered list of nodes obtained after
performing Line 6 of the algorithm for the consensus phase. From a
security point of view, we consider 4 possible events:
\begin{enumerate}
    \item Event ${\mathsf{S}}$: $\bf{s}$ sends a true claim.
    \item  Event $\bar{{\mathsf{S}}}$: The complement of event ${\mathsf{S}}$.
    \item  Event ${\mathsf{N}}$: All nodes in $N_t$ send correct responses.
    \item  Event $\bar{{\mathsf{N}}}$: A non-empty subset of nodes in $N_t$ send incorrect responses.
\end{enumerate}

Notice that there is no question of $\bf{a}$ behaving maliciously
since the claim received by it is for its own benefit. Notice
further that by incorrect response, we mean nodes responding with
$-1$, where the right answer should be either $1$ or $0$. Denote by
${\mathsf{M}}$ the event that $\bf{a}$ decides that $\bf{m}$ is
malicious. We are interested in the four resulting conditional
probabilities. We calculate them sequentially in the following. For
the ease of analysis, we assume that if $\bf{a}$ comes to no
consensus, it will take the claim of $\bf{s}$ as true.

\textbf{Claim 1:} Let $m$ be the number of nodes in $N_t$ that agree
with the claim of $\bf{s}$. Then
$\Pr\left[{\mathsf{M}}|{\mathsf{S}},{\mathsf{N}}\right] = {N_t
\choose m}\sum\nolimits_{i = 1}^m {\frac{{{N_t  - m} \choose
i}}{{2^{N_t  - m} }}}$ when $m < {N_t}/2$ and 1 otherwise.

\textit{Proof:} First assume that $m < {N_t}/2$. The remaining nodes
in $N_t$ will either send $-1$ or $0$ as the response. Event
${\mathsf{M}}$ will be true if the sum of the $-1$'s is less than or
equal to $m$. Assuming $m$ to be fixed, this probability is:
\[
\sum\nolimits_{i = 1}^m {\frac{{{N_t  - m} \choose i}}{{2^{N_t  - m} }}}
\]
Out of $N_t$ nodes, $N_t \choose m$ is the total number of ways in which $m$ nodes can agree with the claim. So the probability is then:
\[
{N_t \choose m}\sum\nolimits_{i = 1}^m {\frac{{{N_t  - m} \choose i}}{{2^{N_t  - m} }}}
\]
The case when $m \ge {N_t}/2$ is obvious.
\begin{flushright}
$\Box $
\end{flushright}

\textbf{Claim 2:} Let $m'$ be the number of nodes in $N_t$ that
send false responses. Let $m$ be the number of nodes in $N_t$ that
agree with the claim of $\bf{s}$. Then
$\Pr\left[{\mathsf{M}}|{\mathsf{S}},{\bar{{\mathsf{N}}}}\right] =
{N_t \choose m - m'}\sum\nolimits_{i = 1}^{m-m'} {\frac{{{N_t  -
m+m'} \choose i}}{{2^{N_t  - m+m'} }}}$ when $m \ge m'$ and 0
otherwise. In particular,
$\Pr\left[{\mathsf{M}}|{\mathsf{S}},{\bar{{\mathsf{N}}}}\right] = 0$
if $m' > {N_t}/2$.

\textit{Proof:} This is analogous to Claim 1, with the  exception
that now $m$ has to be greater than at least $m'$, since otherwise
the sum of responses will be less than 0. Hence we replace $m$ by
$m-m'$ in the probability obtained from Claim 1. The special case
when $m'
> {N_t}/2$ is obvious since then the sum of the responses will
always be less than 0.
\begin{flushright}
$\Box $
\end{flushright}

\textbf{Claim 3:}
$\Pr\left[{\mathsf{M}}|{\bar{{\mathsf{S}}}},{\mathsf{N}}\right] = 1 - \Pr\left[{\mathsf{M}}|{\mathsf{S}},{\mathsf{N}}\right]$.

\textit{Proof:}
Since now the number of nodes that agree with $\bf{s}$ will play an opposite role, the result follows.
\begin{flushright}
$\Box $
\end{flushright}

\textbf{Claim 4:} Let $m'$ be the number of nodes  in $N_t$ that
send false responses. Let $m$ be the number of nodes in $N_t$ that
agree with the claim of $\bf{s}$. Then $
\Pr\left[{\mathsf{M}}|{\bar{{\mathsf{S}}}},{\bar{{\mathsf{N}}}}\right]
= {N_t \choose {m+m'}}\sum\nolimits_{i = 1}^{m+m'} {\frac{{{N_t  -
m-m'} \choose i}}{{2^{N_t  - m-m'} }}}$ when $m+m' < {N_t}/2$ and 1
otherwise.

\textit{Proof:} This is analogous to the proof of Claim 1. Notice
that now there are a total number of $m + m'$ nodes that agree with
$\bf{s}$. Thus we simply replace $m$ by $m + m'$ to complete the
proof.
\begin{flushright}
$\Box $
\end{flushright}

Finally we look at the event when $\bf{a}$ marks $\bf{s}$ as
malicious. This will happen if $\bf{a}$ comes to a consensus
opposite to the claim of $\bf{s}$. Let this event be denoted as
$\mathsf{O}$. We are interested in
$\Pr\left[{\mathsf{O}}|{\mathsf{S}}\right]$ and
$\Pr\left[{\mathsf{O}}|{\bar{{\mathsf{S}}}}\right]$. Let $p = \Pr
\left[ {\mathsf{N}}\right]$. We have the following straightforward
result:

\textbf{Claim 5:}
We have:
\[
\Pr\left[{\mathsf{O}}|{\mathsf{S}}\right] =p(1- \Pr\left[{\mathsf{M}}|{\mathsf{S}},{\mathsf{N}}\right]) + (1-p) (1- \Pr\left[{\mathsf{M}}|{\mathsf{S}},{\bar{{\mathsf{N}}}}\right])
\]
\[
\Pr\left[{\mathsf{O}}|{\bar{{\mathsf{S}}}}\right] =p(1- \Pr\left[{\mathsf{M}}|{\bar{{\mathsf{S}}}},{\mathsf{N}}\right]) + (1-p) (1- \Pr\left[{\mathsf{M}}|{\bar{{\mathsf{S}}}},{\bar{{\mathsf{N}}}}\right])
\]

The results that  we obtain above are an upper bound on the
adversary's limitations. This analysis provides a general
probability method for the determination of certain security metric.

\section{Conclusion and Future Work}
Existing cooperative-based distributed anomaly and intrusion
detection schemes of WSNs do not provide assurance that the
reports/alerts/claims received by the other node(s) were really sent
by the trusted legitimate node(s). Therefore, in this paper we have
proposed the first validation algorithm for trusting anomalies and
intrusion claims. This algorithm uses the concept of an intrusion-aware
reliability parameter that helps to provide adequate reliability at
a modest communication cost.

The proposed work is based  on a few strict assumptions, i.e.,
multiple nodes can sense same anomaly or intrusion. In practice, it
is quite possible that only one node can detect some specific
anomaly or intrusion. Our proposed scheme does not adequately deal with this case. Therefore, more work is needed to make the
proposed scheme further flexible.

\section*{Acknowledgments}
This research was supported by the MKE (Ministry of Knowledge Economy), Korea, under the ITRC
(Information Technology Research Center) support program supervised by the IITA (Institute of Information
Technology Advancement) (IITA-2009-(C1090-0902-0002)) and was supported by the IT R\&D
program of MKE/KEIT, [10032105, Development of Realistic Multiverse Game Engine Technology].
This work also was supported by the Brain Korea 21 projects and Korea Science \& Engineering Foundation
(KOSEF) grant funded by the Korea government (MOST) (No. 2008-1342).


\bibliographystyle{mdpi}
\makeatletter
\renewcommand\@biblabel[1]{#1. }
\makeatother

\vspace{12pt}\noindent \copyright \ 2009 by the authors; licensee Molecular Diversity Preservation International, Basel, Switzerland.
This article is an open-access article distributed under the terms and conditions of the Creative
Commons Attribution license (http://creativecommons.org/licenses/by/3.0/).

\end{document}